\documentstyle[multicol,epsfig,pre,aps,amssymb]{revtex}

\begin{document}

\title{Evolution of turbulent spots in a parallel shear flow }
\author{\sc J\"org~Schumacher and Bruno~Eckhardt}
\address{Fachbereich Physik, Philipps-Universit\"at Marburg,
D-35032 Marburg, Germany}
\date{\today}
\maketitle

\begin{abstract}
The evolution of turbulent spots in a parallel shear flow is studied by means of
full three-dimensional numerical simulations.  The flow is bounded by free
surfaces and driven by a volume force.  Three regions in the spanwise spot
cross-section can be identified:  a turbulent interior, an interface layer with
prominent streamwise streaks and vortices and a laminar exterior region with a
large scale flow induced by the presence of the spot.  The lift-up of streamwise
streaks which is caused by non-normal amplification is clearly detected in the
region adjacent to the spot interface.  The spot can be characterized by an
exponentially decaying front that moves with a speed different from that of the
cross-stream outflow or the spanwise phase velocity of the streamwise roll
pattern.  Growth of the spots seems to be intimately connected to the large scale
outside flow, for a turbulent ribbon extending across the box in downstream
direction does not show the large scale flow and does not grow.  Quantitatively,
the large scale flow induces a linear instability in the neighborhood of the
spot, but the associated front velocity is too small to explain the spot
spreading.
\end{abstract}

\begin{multicols}{2}
\section{Introduction}
The transition to turbulence in spatially extended systems does not necessarily
take place in all points simultaneously but can be preceeded by the formation of
localized structures that grow to eventually fill space.  Already the first
experiments by Reynolds in pipe flow showed the formation of turbulent spots and
slugs \cite{Rey83,Cha73}.  In Couette-Taylor flow between counterrotating
cylinders turbulence can be confined to propagating spirals \cite{Col65,Heg89}.
Localized turbulence has 
also been observed in plane Couette flow \cite{Til92} where the
fraction of space filled with turbulent flow has been used as a measure to define
the transition to turbulence \cite{Dav92,Bo98,Man00}.  Besides these transitional
phenomena localized turbulent spots can also be observed in high Reynolds number
boundary layers \cite{Gad81}.

It is tempting to connect both the localization of the spots, i.e.  the
coexistence of a laminar and turbulent phase of the shear flow, and the
propagation of the sharp boundaries, i.e.  a front-like structure, to phenomena
studied in considerable detail within amplitude models \cite{Cro92}.  Indeed,
some models show qualitatively similar behavior.  There are, however, several
problems that raise questions about the applicability of such models.  For
instance, they are not derived from the Navier-Stokes equation and the extend to
which they reflect the hydrodynamical processes and interactions remains open.
Furthermore, amplitude equations work best if they can be applied in a situation
of linear instability and small amplitudes \cite{Pom91}, such as the onset of
Rayleigh-B\'{e}nard convection near the critical point \cite{Cro80}.  
But many of the turbulent spots arise in
shear flows that are linearly stable, at least in the Reynolds number region of
interest here.  Such behavior can be captured in higher-order Ginzburg-Landau
models \cite{Saa92}, but the required large amplitudes complicate a quantitative
comparison.  Moreover, investigations of plane Couette flow show that the
turbulent state is not stable but can decay spontaneously for lower Reynolds
number values \cite{Schm97,Schm99,Gro99}.

It is our aim here to analyze the evolution of turbulent spots in parallel shear
flows, in particular their spanwise spreading.  Our flow has free-slip boundary
conditions and is driven by a volume force.  Despite the change in boundary
conditions we observe features similar to those in experiments on
plane Couette flow with rigid boundary conditions and a linear shear profile:
this supports the expectation that there are perhaps universal aspects.  The
model is moreover well suited for high resolution direct numerical simulations
with a Fourier-pseudospectral method and allows for a detailed investigation of
the dynamics in the transitional region.  In particular, we focus on the
characterization of the front which separates the laminar and the turbulent
region, on the mechanism by which it propagates, on the Reynolds number
dependence of the front speed and on the large scale flow in the laminar
surrounding of the spot.  As we will discuss in more detail in the appropriate
sections these aspects complement previous numerical and experimental
investigations in wall bounded shear flows
\cite{Hen87,Hen89,Lun91,Lun93,Til95,Dav95,Mal95,Heg96}.

The paper is arranged as follows.  After introducing the physical model and the
numerical procedures in Sec.~II we discuss in Sec.~III the hydrodynamics of the
spreading mechanism in some detail.  The properties of the tail of the envelope,
such as spatial decay and spreading velocity, are discussed and three different
regimes of the spreading process are identified.  In Sec.~IV we discuss the
results and give a brief outlook.

\section{The model}
\label{sec_basic}
The system we consider here is a shear flow between parallel free-slip surfaces
and driven by a volume force.  In the streamwise and spanwise direction periodic
boundary conditions are applied; in the normal direction the normal velocity
component vanishes in the two bounding surfaces.  With lengths measured in units
of $d/2$ (half the gap width) the periodicities in streamwise and spanwise
directions are both $80$.  The volume force with a sinusoidal dependence in
normal direction gives rise to a laminar profile with velocities $\pm U_0$ at the
surfaces.  The Reynolds number is defined as $Re=U_0 d/(2\nu)$.  In these units
the incompressible Navier--Stokes equation for a velocity field ${\bf u}({\bf
x},t)$ becomes
\begin{eqnarray}
\label{nseq}
\frac{\partial{\bf u}}{\partial t}+({\bf u}\cdot{\bf \nabla}){\bf u}
&=&-{\bf \nabla} p+\frac{1}{Re}{\bf \nabla}^2{\bf u}+{\bf f}\;,\\
{\bf \nabla}\cdot{\bf u}&=&0\;
\end{eqnarray}
$p({\bf x},t)$ denotes the
pressure  and ${\bf f}({\bf x},t)$ the external volume force specified
below.  

\begin{figure}
\begin{center}
\epsfig{file=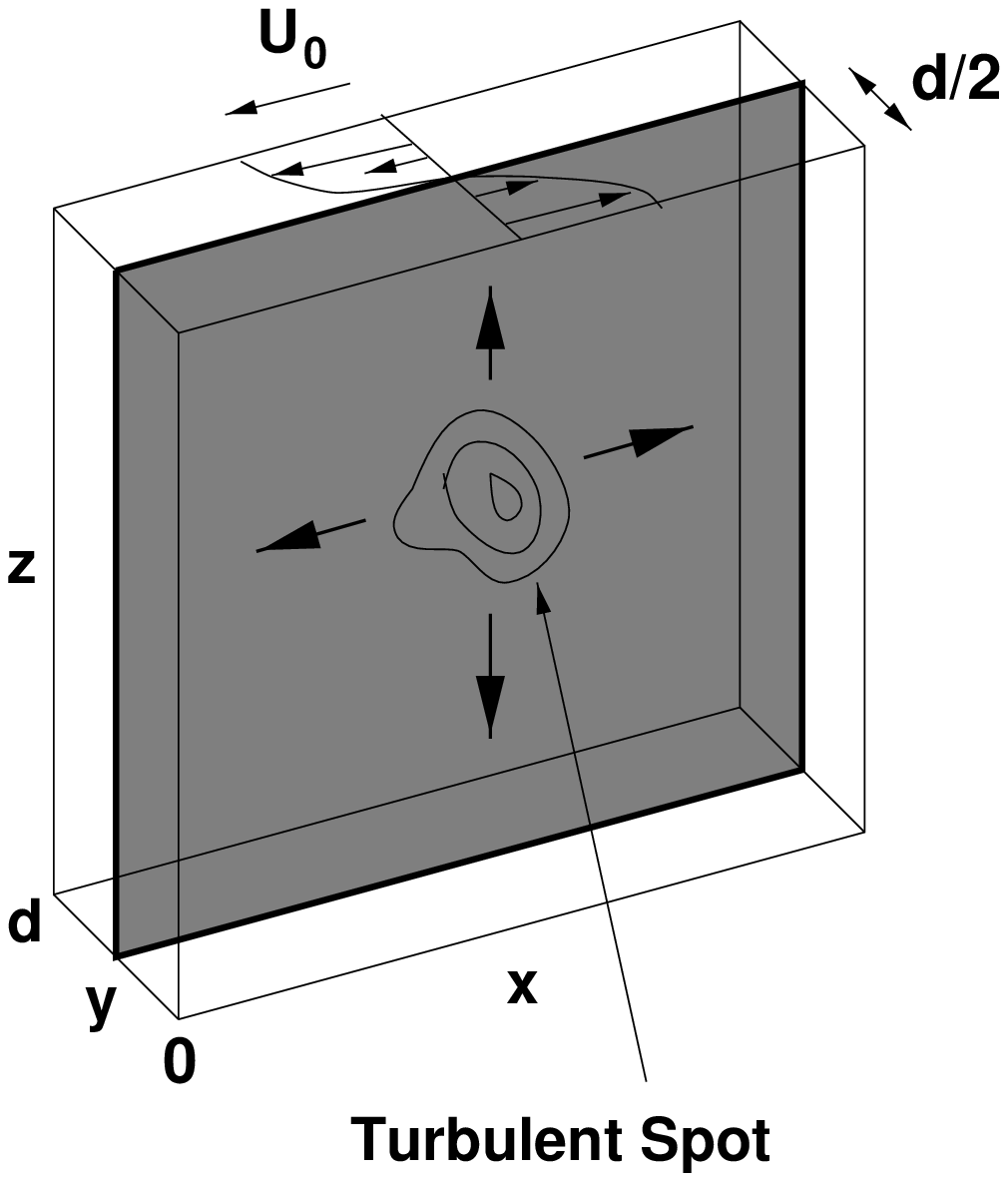,width=7cm}
\end{center}
\end{figure}

\noindent
FIG. 1. Geometry of the flow.  The $x$ axis points in the streamwise, 
$y$ in the wall-normal, and $z$ in the
spanwise direction.  The central plane at $y=1$ in
which the spreading is analysed is shaded gray.

\vspace{0.5cm}
\noindent
Figure~1 shows the Cartesian coordinate system
we use, with $x$ pointing in the 
streamwise, $y$ in the wall--normal and $z$ in the spanwise directions. 
The fluid volume is confined to $0\le y\le 2$ with boundary conditions 
\begin{equation}
\label{bc_stress-free}
u_y=\frac{\partial u_x}{\partial y}=\frac{\partial u_z}{\partial y}=0
\hspace{1em} 
\text{at } y=0 \mbox{\ and \ } 2
\end{equation}
at the surfaces and periodic boundary conditions in downstream
and spanwise directions.

The shear flow is driven by a volume force ${\bf f}=\pi^2/(4Re) \cos(\pi
y/2)\,{\bf e}_x$ acting in the $x$-direction, which in the laminar regime
sustains a flow ${\bf U}_0=\cos(\pi y/2)\,{\bf e}_x$.  The velocity field ${\bf
u}({\bf x},t)$ is decomposed into this laminar flow ${\bf U}_0$ and a turbulent
part ${\bf v}({\bf x},t)$.  As shown already by Tollmien, ${\bf U}_0$ is linearly
stable, thus demonstrating that Fj{\o}rtoft's theorem is a necessary but not
sufficient condition for the transition to turbulence\cite{Dra81}.  Nevertheless,
for sufficiently large driving the flow shows a transition to turbulence.

The free-slip or stress-free boundary conditions have the advantage that the flow
can be represented completely by Fourier modes so that robust pseudospectral
techniques based on a 2/3-rule dealiazing and an adaptive Runge-Kutta scheme for
advancing in time can be used\cite{Can88,See96}.  To account for boundary
conditions the flow is represented by the Fourier sums
\begin{eqnarray}
\label{fansatz_a}
u_x({\bf x},t)&=&\sum_{\bf k} u_{x{\bf k}}(t)\cos(k_y y)\exp[i(k_x x+k_z z)]\;,\\
u_y({\bf x},t)&=&\sum_{\bf k} u_{y{\bf k}}(t)\sin(k_y y)\exp[i(k_x x+k_z z)]\;,\\
\label{fansatz_e}
u_z({\bf x},t)&=&\sum_{\bf k} u_{z{\bf k}}(t)\cos(k_y y)\exp[i(k_x x+k_z z)]\;.
\end{eqnarray}
with wavenumbers 
\begin{eqnarray}
\label{wellenzahlen}
k_y&=&0,\frac{\pi}{2}, \pi, \dots \frac{ N_y\pi}{2}\;,\\
k_x&=&0, \pm \frac{2\pi}{L_x}, \pm 2\frac{2\pi}{L_x},\dots \pm
\frac{N_x}{2}\frac{2\pi}{L_x} \;,\\ 
k_z&=&0, \pm \frac{2\pi}{L_z}, \pm 2\frac{2\pi}{L_z},\dots \pm
\frac{N_z}{2}\frac{2\pi}{L_z}\;.
\end{eqnarray}
In \cite{Wal97} and \cite{Man00}, respectively, low-dimensional models for the
transition to turbulence in plane shear flows with stress-free boundary
conditions were discussed.  Their basic flow has the form $U_{0x}(y)\sim\sin(\pi
y/2)$ and is confined to an interval $y\in [-1,1]$.  Both expansions can thus be
related by a shift in the interval, but we prefer
(\ref{fansatz_a})-(\ref{fansatz_e}) as it has the more compact representation in
sines and cosines and is easier to implement numerically.  The spectral
resolution for all runs with evolving spots was $N_x\times N_y\times
N_z=256\times 33\times 512$.  The initial localized perturbation is a 
poloidal vortex of the form
\end{multicols}
\begin{eqnarray}
\label{disturbance}
{\bf v}({\bf x},t=0)={\bf \nabla\times\nabla\times}
A\exp[-a_x^2(x-x_0)^2-a_y^2(y-y_0)^2-a_z^2(z-z_0)^2]{\bf e}_y\;,
\end{eqnarray}
\begin{multicols}{2}
positioned slightly off center in order to avoid spurious effects
due to accidental symmetries.
This initial condition is a model for the flow induced in 
experiments where a small transverse jet penetrates the laminar 
shear profile in wall-normal direction \cite{Til92,Dav92,Bo98}.

\begin{figure}
\begin{center}
\epsfig{file=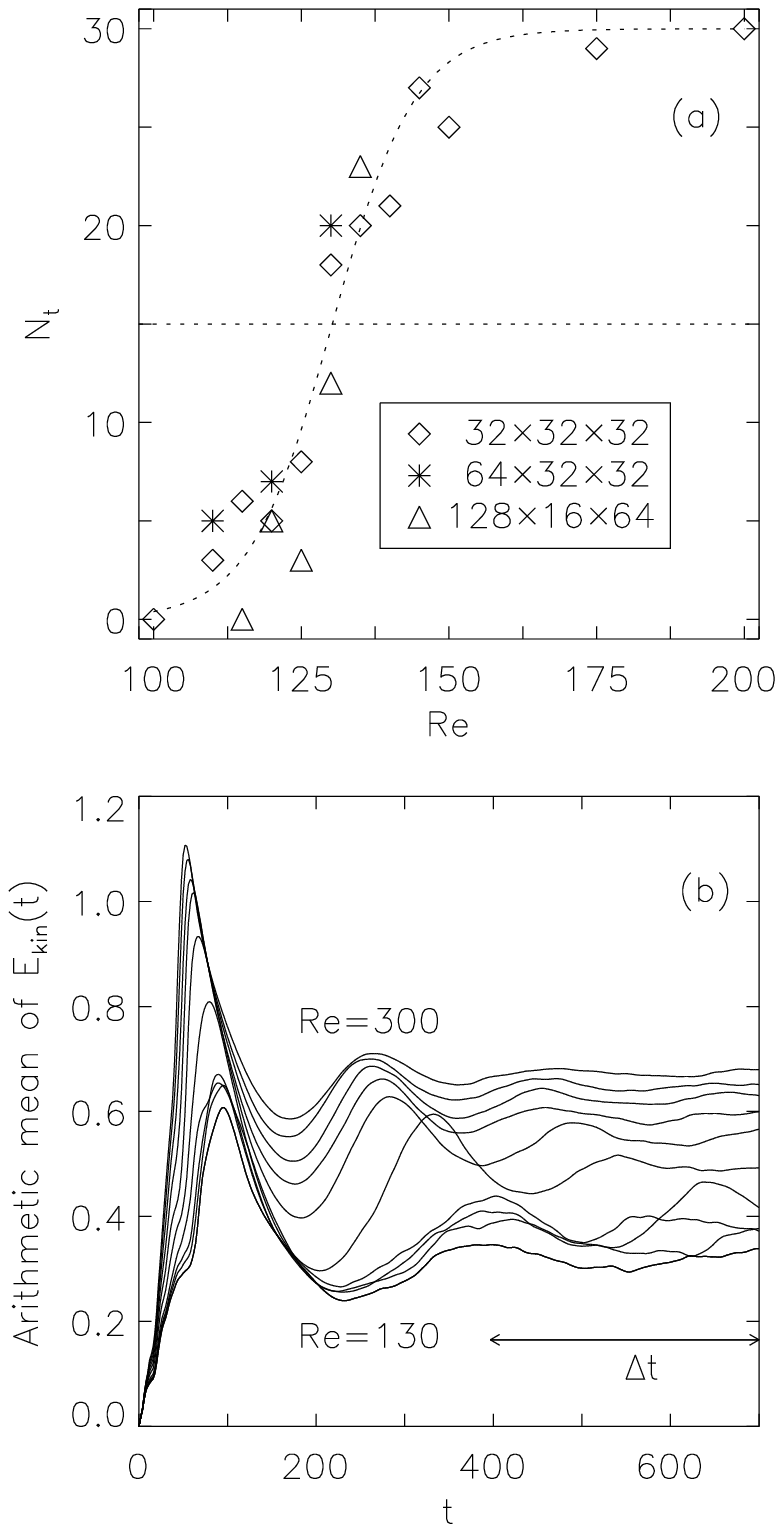,width=6cm}
\end{center}
\end{figure}

\noindent
FIG. 2. Statistical analysis of the transition to turbulence.
Panel (a) shows the number $N_t$ of initial conditions that lead to persistent
turbulence for different Reynolds
numbers and for several spectral resolutions. Thirty trajectories were
run for every Reynolds number. The dashed line
is a two parameter least square fit $N_t(Re)=15\times\tanh[(Re-A_0)/A_1]+15$.
Panel (b) shows the temporal relaxation of the turbulent kinetic energy,
averaged over all trajectories that became turbulent. The double headed arrow
marks the interval of temporal averaging used for Eq.~(\ref{level}).

\vspace{0.5cm}
\noindent
For the analysis in the following sections we need the Reynolds number above
which a transition to turbulence occurs.  Because of the free-slip boundary
conditions at the surface it can be expected to be below the one for rigid walls.
But as in that case the transition is strongly intermittent and the best approach
to a definition of a critical Reynolds number uses a statistical analysis of run
time experiments with different initial conditions \cite{Schm99a}.  At each value
of $Re$ we run thirty trajectories starting from states with slightly different
amplitudes $A$. The different initial conditions were obtained
by switching on a field (\ref{disturbance}) with 
time-dependent amplitude $A(t)=a_0\sin^2(\pi t/2)$ for 
$t\in[0,2]$, where the factor $a_0$ was increased from 1.248 to 1.326 in steps 
of 0.0026. 
Since the aim is to obtain
information on the bulk properties, we took a smaller box with aspect ratio
$L_x:L_y:L_z=40:2:20$ (in units of half the gap width), and a lower spectral
resolution.  As in the direct numerical simulations of plane Couette flow
\cite{Schm99a}, two types of dynamics can be identified:  in one case the flow
builts up and the energy relaxes with oscillations to the turbulent state, i.e.
$E_{kin}\ne 0$, whereas
in the other case the state decays towards the laminar profile, so that finally
$E_{kin}$ becomes negligible.  The number $N_t$ of initial conditions that became
turbulent vs. Reynolds number is shown in panel (a) of
Fig.~2.  For $Re\gtrsim 200$ all samples relax to the turbulent state and
the turbulent energy increases linearly with Reynolds number, giving rise to the
relation (\ref{level}). As in the case of plane Couette flow, the relaxation 
to the turbulent state is oscillatory (Fig.~2b). We conclude from 
these studies that more than half the initial conditions will become turbulent
for Reynolds numbers of $130\pm 5$, which, as expected, is lower than the value
of  $320\pm 10$ 
for rigid boundary conditions \cite{Schm99a}. Below this Reynolds number
most turbulence is transient and a turbulent region can disappear
by erosion from within. Thus, the kind of spot spreading phenomenon
we are interested can only occur for Reynolds numbers above this
value.

\section{Spreading of the turbulent spot}
After a short transient of about $5$ time units the inital perturbation
which was localized in diameter to about 4 in half gap width units
develops streamwise streaks and vortices and starts to expand.
To highlight the turbulent deviations 
about the laminar flow a contour plot of the 
downstream velocity averaged over half a box height,
\begin{equation}
\bar{v}_x(x,z,t)=\int_0^1 \,v_x(x,y,z,t)\,\mbox{d}\,y\;,
\end{equation} 
is shown in Fig.~3 (left panel).  The elongated streamwise streaks with
alternating flow direction stand out above the 
background flow.  The cut at $x=L_x/2=40$ for the
streamwise turbulent velocity itself underlines the existence of the streamwise
streaks (right panel).  In the streamwise direction the spot advances more or
less stochastically, with the unpredictable appearance of turbulent bursts which
are then advected by the laminar profile.  This causes a strongly fragmented
spot interface.  In the spanwise direction it advances more steadily with a
regular interface and it is this direction we focus on in the following
analysis. 
\end{multicols}
\begin{figure}
\begin{center}
\epsfig{file=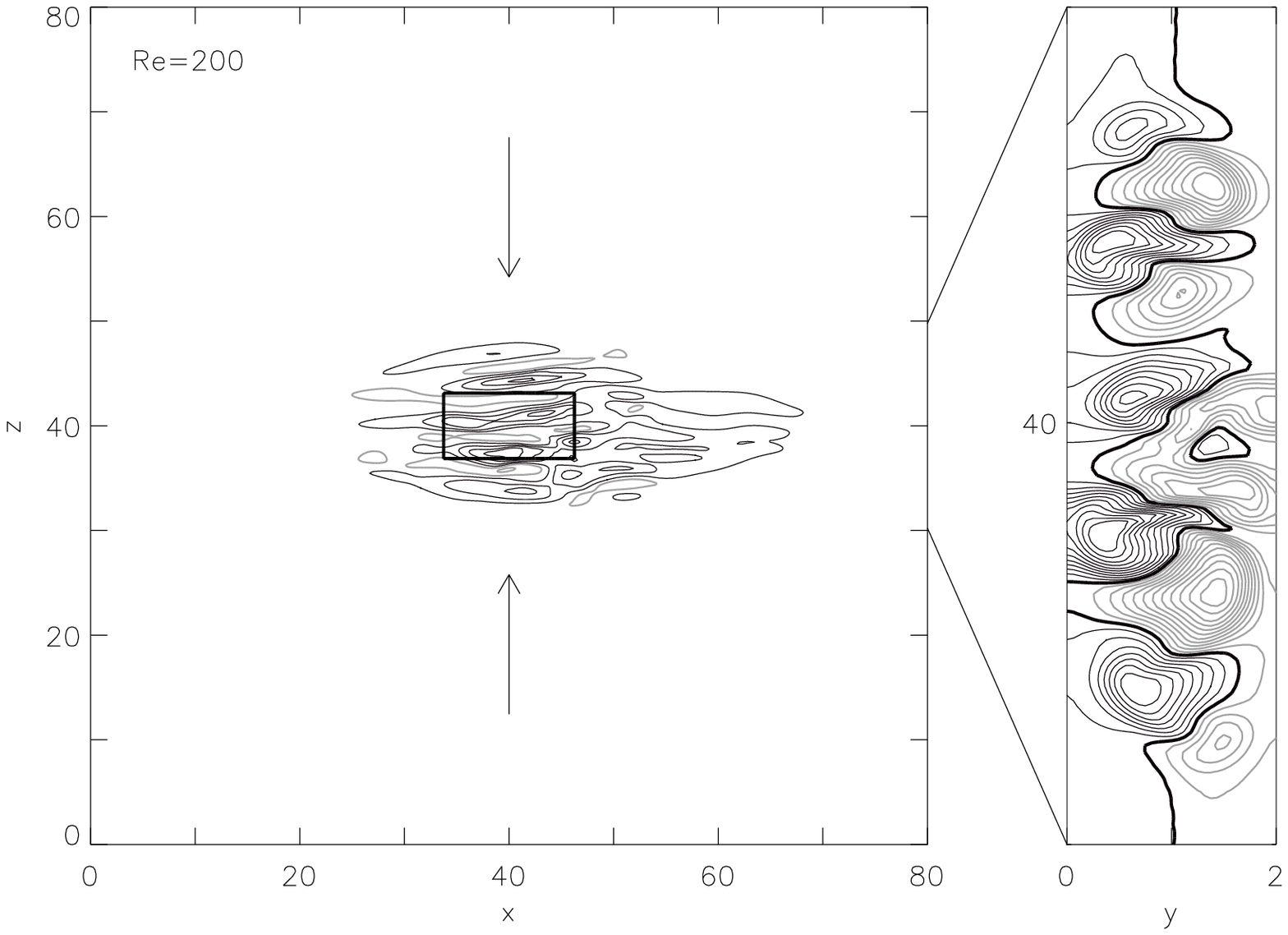,width=13cm}
\end{center}
\end{figure}

\noindent
FIG. 3. Streamwise streaks in an expanding spot for $Re=200$ at $t=39$.  Left:
Contours of $\bar{v}_x(x,z)$.  The rectangular box in the spot center
marks the lateral extension of
the volume used for the analysis of internal turbulent fluctuations.  Right:
Magnification of the wall-normal--spanwise plane at $x=L_x/2$ (marked by arrows
in the left panel).  Corresponding contours of $v_x(x=L_x/2,y,z)$ are plotted
using the same linestyle:  gray lines denote negative values, black lines
positive ones, and the heavy solid line is $v_x=0$.


\begin{multicols}{2}
Cross sections of the local turbulent energy ${\bf v}^2$ taken
in the middle of the cell at $x=L_x/2$ and $y=1$ for fixed time 
indicate three different flow regimes: a turbulent interior,
a laminar exterior and a narrow transitional region with
large velocity amplitudes and a rather regular spatial structure
(see Fig.~4). We will discuss these regions in turn.

\subsection{Turbulent interior and wave propagation at the spot interface}
\label{interior_chap}

The turbulent fluctuations in the interior were investigated in detail for five
values of the Reynolds number, as shown in Tab.~\ref{Tab1}.  The fluctuations, in
units of $U_0^2$, were obtained by averaging over a box $V_c$ of size 
$l_x:l_y:l_z=12:2:6$ (see the left panel of Fig.~3) in the
center of the spot according to
\begin{equation}
\langle v_i^2\rangle (t_0) =\frac{1}{V_c}\int_{V_c} \,v_i^2(x,y,z,t_0)\,\mbox{d}\,V\;\;\;
\text{for}\;\;\; i=x, y, z\,.
\end{equation} 
\begin{figure}
\begin{center}
\epsfig{file=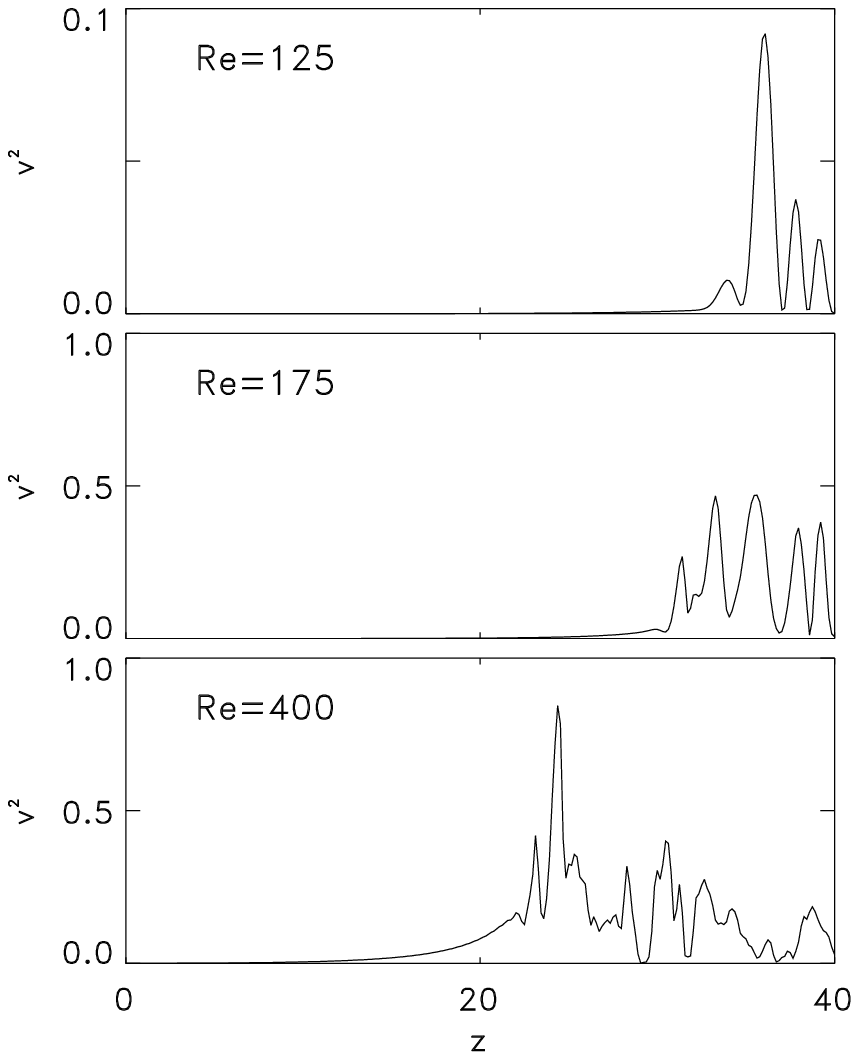,width=7cm}
\end{center}
\end{figure}

\noindent
FIG. 4. Envelopes of the turbulent kinetic energy along the spanwise axis
for three different Reynolds numbers at $t=59$ after inducing the perturbation.

\end{multicols}
\begin{figure}
\begin{center}
\epsfig{file=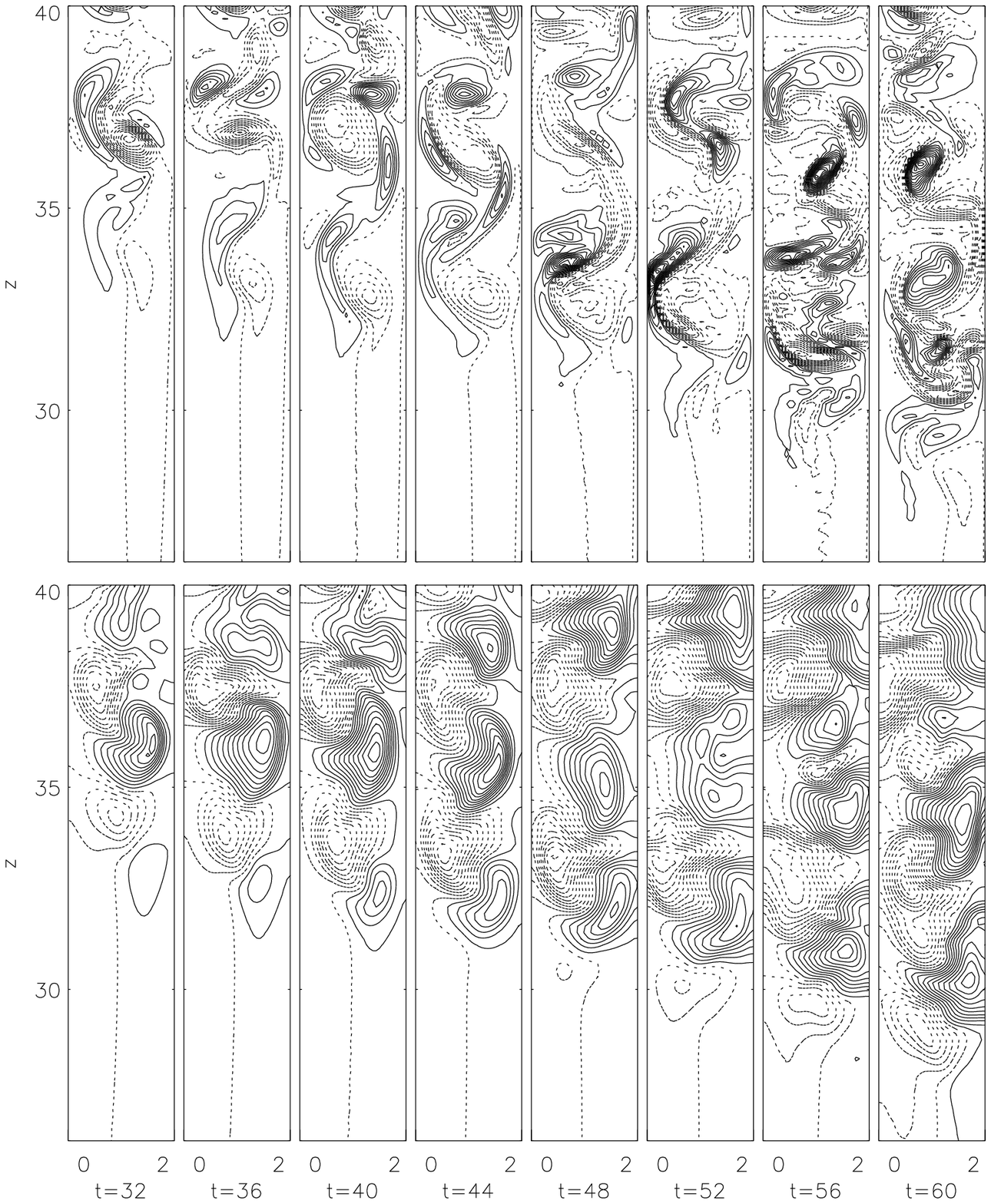,width=13cm}
\end{center}
\end{figure}

\noindent
FIG. 5. Snapshots of the streamwise vorticity $\omega_x$ (upper row)
and the streamwise velocity $v_x$ (lower row). The data 
are for $Re=200$ at $x=L_x/2=40$. Dotted lines denote positive
contours and solid lines negative ones. Horizontal axis denotes 
wall-normal direction in all panels.
\begin{multicols}{2}

For a Reynolds number of about 150, where the spot decays, the fluctuations
decrease with time.  For higher Reynolds numbers they first increase until a time
of about 30 and then stay constant, within statistical fluctuations.  The values
quoted in the table are the temporal averages taken between $t=30$ and 50.
As expected, the streamwise fluctuations are largest but smaller than
the fluctuations in the interface region (see Fig.~4).

The spatial modulations near the interface are due to
elongated streamwise structures, so-called streamwise streaks, which are
not stable but travel slowly in spanwise direction with a phase velocity $v_w$.
The occurence of such (oblique) waves was noted in several previous
investigations, mainly for plane Poiseuille flow, and has been connected to an
inflectional instability of the combined shear and cross flow velocity field 
\cite{Hen87,Til95}. A linear stability analysis of the combined profile for
plane Poiseuille and plane Couette flow gives a range of critical wavenumbers 
(see \cite{Hen89,Lun93} and below). 

The spreading of the spot in the spanwise direction is further documented in
Fig.~5 for $Re=200$.  The streamwise vorticity component
$\omega_x=\partial_y v_z-\partial_z v_y$ is shown in the upper panels.  The
growth of a pair of new counterrotating vortices (panels No.  2,3, and 4 of
Fig.~5) can be followed for all eight snapshots.  The corresponding
streamwise streak which is lifted up by the non-normal amplification can also be
identified in the cross-section of the streamwise velocity $v_x$ (see the lower
panels).  Although there is a certain discreteness in the growth of new
streamwise rolls and pairs of streamwise vortices, the front advances steadily in
time without any disruptions as will be shown later in Sec.~\ref{front_chap}.
\end{multicols} 
\begin{table}
\begin{tabular}{ccccccc}
$Re$   & $v_w$ & $v_F$ 
& $\sqrt{\langle v^2\rangle}$ 
& $\sqrt{\langle v_x^2\rangle}$ 
& $\sqrt{\langle v_y^2\rangle}$ 
& $\sqrt{\langle v_z^2\rangle}$\\
\hline
$150$  & 0.02 & 0.09 & 0.19 & 0.18 & 0.02 & 0.05\\ 
$200$  & 0.08 & 0.42 & 0.41 & 0.38 & 0.09 & 0.12\\ 
$250$  & 0.14 & 0.56 & 0.50 & 0.45 & 0.13 & 0.19\\ 
$300$  & 0.17 & 0.63 & 0.61 & 0.56 & 0.15 & 0.19\\
$350$  & 0.16 & 0.64 & 0.57 & 0.53 & 0.13 & 0.19\\
\end{tabular}
\vspace{0.5cm}
\caption{Characteristic velocities of the spreading spot in units of $U_0$
for the several Reynolds
numbers: the phase velocity $v_w$, the front velocity $v_F$, and the root
mean square velocities of the three components $v_i$ with $i=x, y$ and $z$.
The phase velocity was determined separately for the maxima and minima
of $v_x$ closest to the boundary and their arithmetic mean is listed.
The turbulent fluctuations taken over a small volume in the spot center
were in addition averaged over time between $t=30$ and 50. Third column is 
the root mean square velocity, 
$(\langle v_x^2\rangle+\langle v_y^2\rangle+\langle v_z^2\rangle)^{1/2}$.}
\label{Tab1}
\end{table}
\begin{multicols}{2}
In Table~\ref{Tab1} we have included the results for the phase velocity $v_w$ of
the streaks.  The wavelength was also determined as the distance between two
streaks with the same sign, i.e., the spatial distance between two maxima or
minima of $v_x$, measured in units of $d/2$.  We found values between 2.4 and 3.4
with a tendency towards smaller values for higher $Re$.  Velocities were
determined by monitoring the motion of these maxima and minima in $v_x$.  Only
extrema closest to the interface were included and measurements were limited to
times between $t=32$ and $t=60$ for all five data sets.  We find that the phase
speed increases with Reynolds number. 

\subsection{The large scale flow outside the spot}
Near the spanwise centerline across the spot we find a strong outward
pointing flow. It varies with height but is not compensated by
an inflow on any level. Incompressibility thus demands a compensating
inflow in other parts of the spot. 
The magnitude of this large scale flow decreases rather rapidly with 
distance from the spot and accounts almost completely for the deviations from
the laminar profile. 
In order to highlight the flow pattern we show in 
Fig.~6 the directional field, i.e. $\overline{{\bf
v}}/|\overline{{\bf v}}|$, where the overbar indicates
an average in the normal direction. The flow 
has quadrupolar characteristics, with outflow in the spanwise direction
and inflow in the streamwise direction.

The outward velocity does not coincide with the phase speed of the
waves.  But
small as the large scale flow may be, it has profound consequences for the
spreading of the spot.  Consider, for instance, the case of a ribbon spanning
the periodicity box in streamwise direction but localized in spanwise direction:
no such quadrupolar flow can form.  And indeed, the ribbon does not spread!
(see the dashed line in panel (a) of Fig.~9).
\begin{figure}
\begin{center}
\epsfig{file=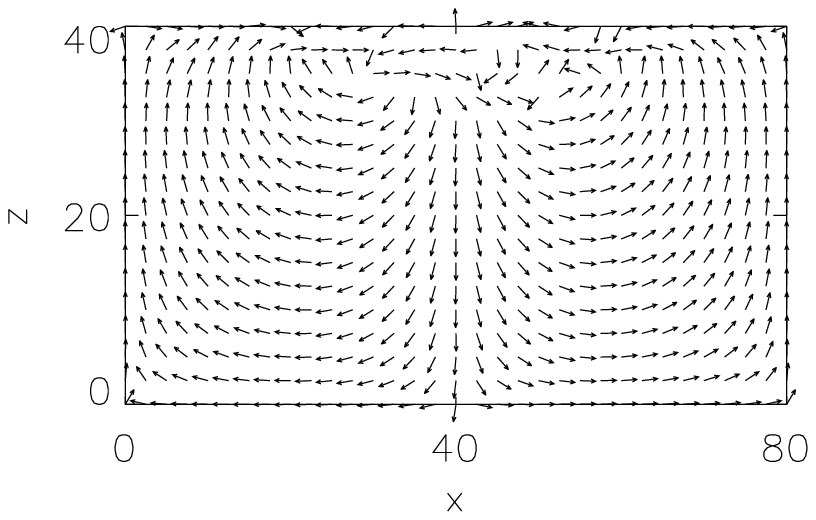,width=6cm}
\end{center}
\end{figure}

\noindent
FIG. 6. Large scale flow outside the spot. To emphasize the topology
of the flow field only the direction $\overline{{\bf v}}/|\overline{{\bf v}}|$,
averaged in wall-normal direction $x$, is shown. The domain 
is the lower half of the full integration domain, with the
center of the spot in the middle of the upper boundary.
The flow is for a Reynolds number of $Re=200$ at a time $t=39$.


\subsection{The propagating front}
\label{front_chap}
Outside the spot and into the laminar regime one notes a gradual decrease 
in local turbulent energy ${\bf v}^2({\bf x},t)$. 
 
Quantitatively, the energy density decays exponentially, as demonstrated in
Fig.~7, where segments of the turbulent spot envelopes of the lower
front (moving toward $z=0$) between $t=34$ and $t=50$ are plotted.  The rate of
spatial decay $\lambda$ grows slowly for Reynolds numbers $Re\ge 200$.  Fitting
to each of 11 snapshots separately an exponential profile
$\sim\exp(-z/\lambda)$ gives decay rates $\lambda$ between $0.9$ and
$1.2$. In Fig.~8 the arithmetic mean and the
corresponding error bars are shown for $Re\ge 175$.  
For smaller Reynolds numbers deviations from an exponential envelope
are larger, resulting in larger variations and uncertainties in the spatial
decay rates. 
This exponentially decaying envelope is not much influenced by the turbulent
fluctuations inside the spot and advances more or less steadily into the laminar
region.  Monitoring the position of a certain turbulent intensity threshold
allows to extract a velocity $v_F$ that is independent of the selected threshold
along the tail.  However, since the turbulent intensity increases with Reynolds
number it is advisable for numerical reasons to also adjust the threshold.  We
choose to increase the level linearly with $Re$,
\begin{equation}
{\bf v}^2=7.5\cdot 10^{-3}\times\,[a_1+a_2(Re-Re_0)]\,,
\label{level}
\end{equation} 
\begin{figure}
\begin{center}
\epsfig{file=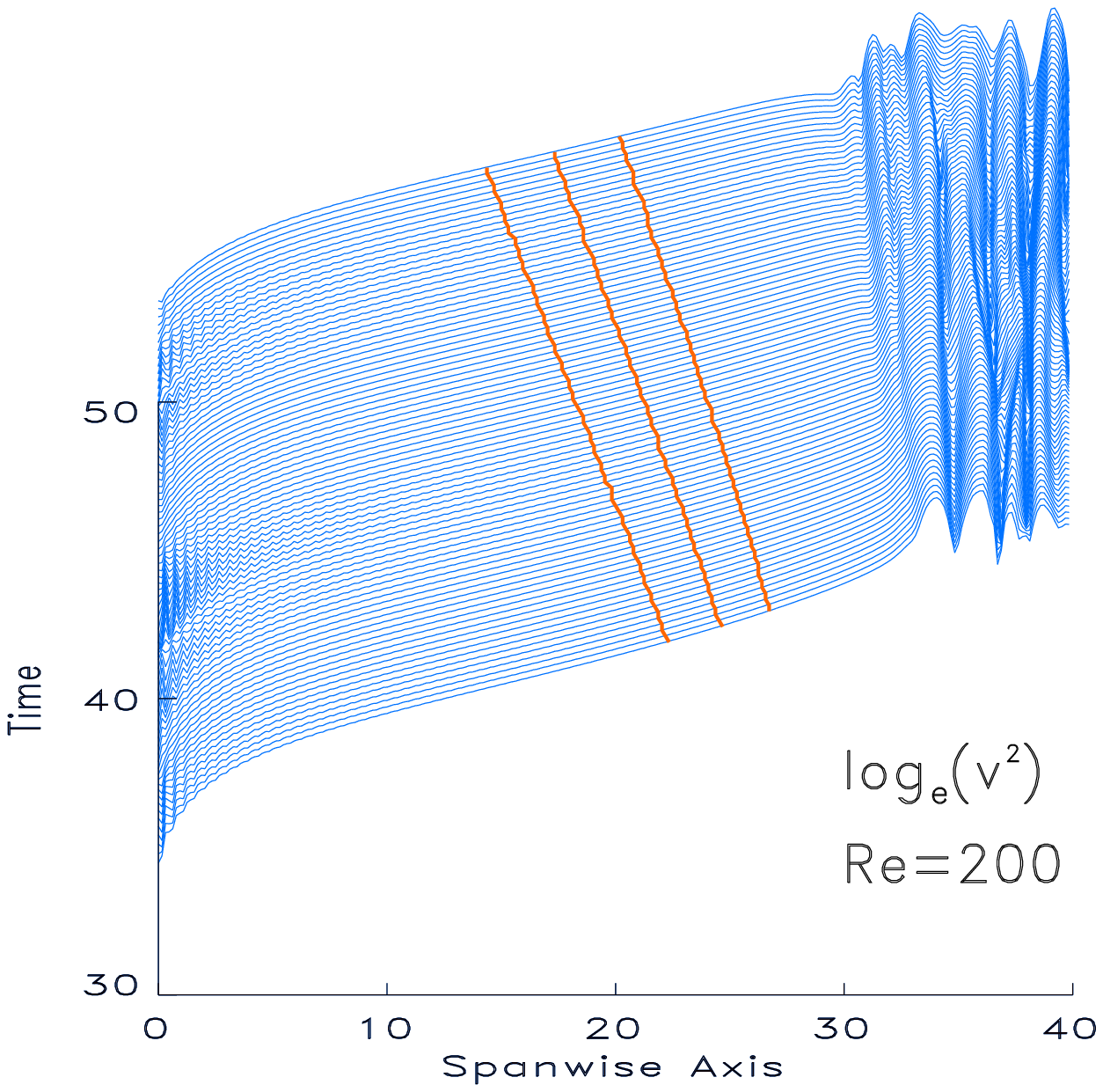,width=6cm}
\end{center}
\end{figure}

\noindent
FIG. 7. The exponential decay towards the laminar profile outside
the turbulent spot at $Re=200$. Shown are the profiles of the
turbulent kinetic energy along the lower half of the center line
on a semi-logarithmic scale. Curves for different 
times in $[34, 50]$ are vertically displaced and overlayed.
The lines across the plot indicate the motion of the levels at which
the front velocity was determined.
\begin{figure}
\begin{center}
\epsfig{file=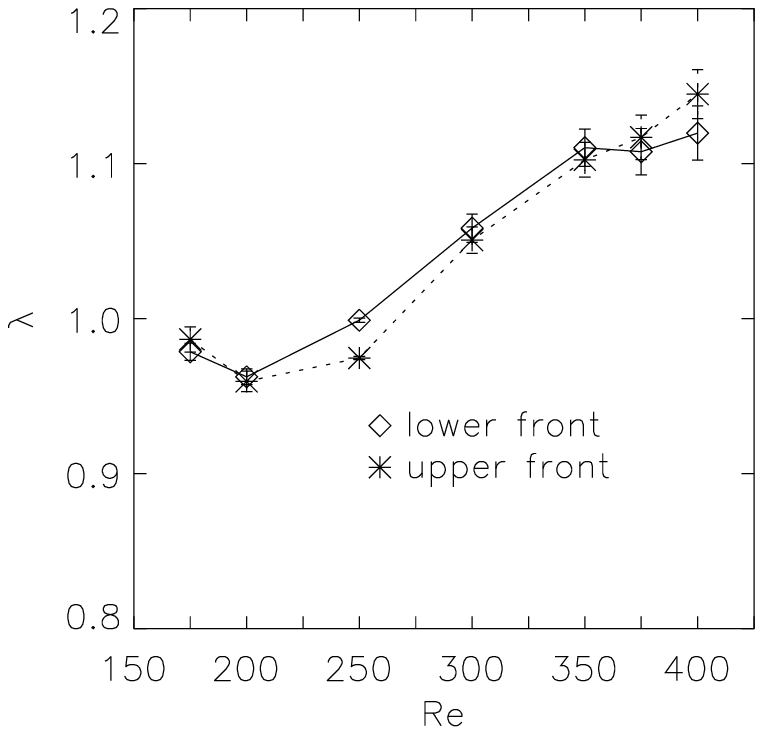,width=6cm}
\end{center}
\end{figure}

\noindent
FIG. 8. Exponential decay rate $\lambda$ of the envelope.
The data are mean values taken from front envelopes as
shown in Fig.~7.

\noindent
where the coefficients
$a_1=0.143$, $a_2=2.6\cdot 10^{-4}$, and $Re_0=200$ follow
from a linear fit to the mean kinetic energy 
$E_{kin}(t)=(1/2V)\int_V {\bf v}({\bf x},t)^2\,\mbox{d}\,V$ 
as a function of $Re$.
This variation of mean kinetic energy was determined alongside with the
statistical analysis needed to determine the Reynolds number 
for the transition to the turbulent state (see section~\ref{sec_basic}).

\begin{figure}
\begin{center}
\epsfig{file=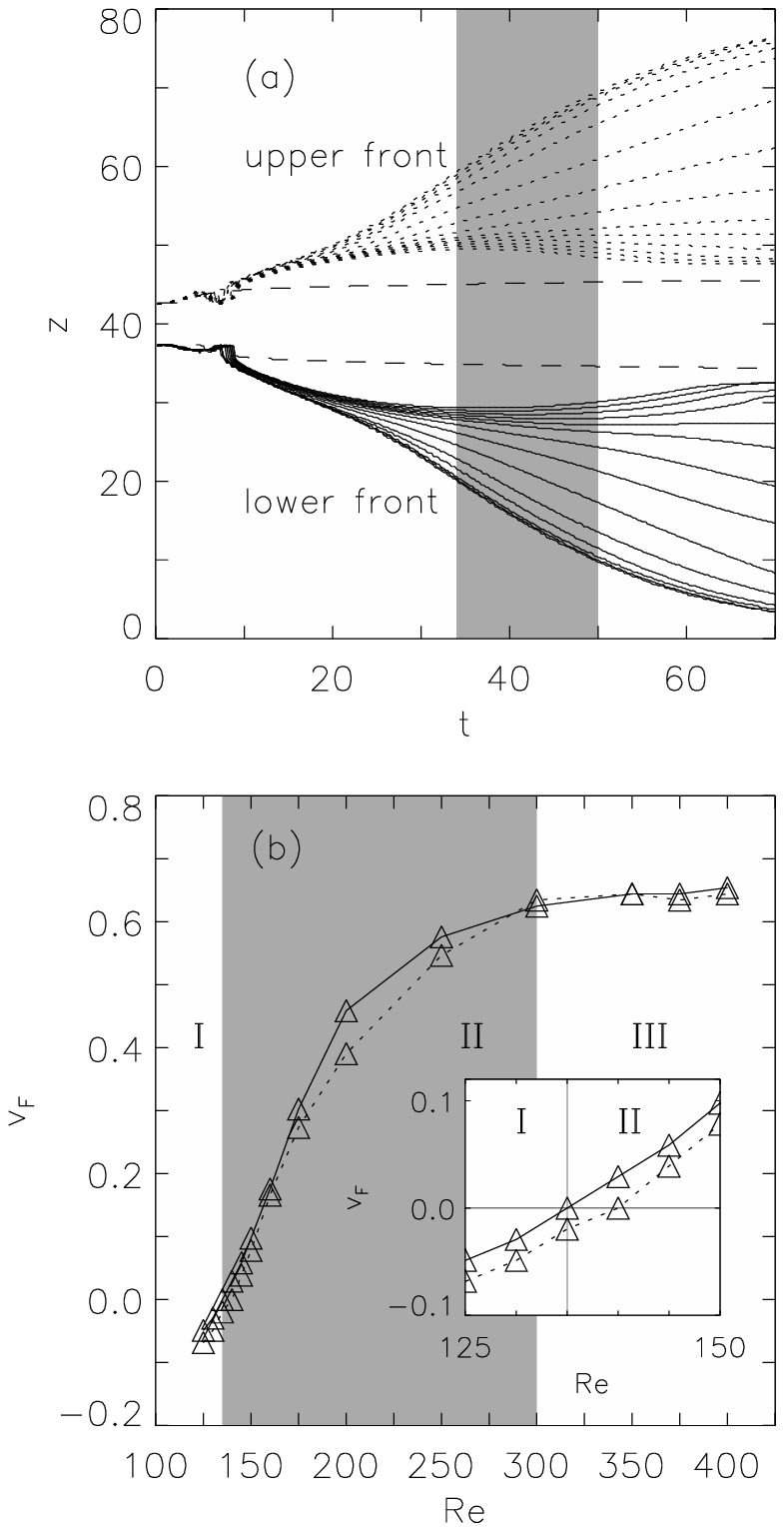,width=6cm}
\end{center}
\end{figure}

\noindent
FIG. 9. Reynolds number dependence of the front velocity.
(a):  Position of the extracted level ${\bf v}^2$ of the envelope along 
the spanwise centerline $(x=40,y=0,z)$ vs. time $t$ for both spanwise fronts. The
outer most curves for the lower and upper front are taken at $Re=400$, the
inner ones at $Re=125$. The dashed lines correspond to a run at $Re=200$
with a turbulent ribbon that extended across the box in
downstream direction. (b):  Corresponding front velocities $v_F$ 
(absolute values)
extracted within
the shaded time interval in the panel (a).  The inset highlights 
the crossover to the shrinking regime around $Re\simeq 135$.

\vspace{0.5cm}
\noindent
The front speed thus determined is shown in panel (b) of Fig.~9.  It
increases for Reynolds numbers between 135 and 200 and saturates for Reynolds
numbers above 300.  Both regimes were also found in plane Couette flow
experiments with rigid walls.  Dauchot and Daviaud \cite{Dav95} observed an
increasing spreading rate for $Re$ between 370 and 450.  Tillmark and Alfredsson
found a constant spreading rate for $Re\gtrsim 500$ \cite{Til92}.  In our
simulations we can cover both ranges.  Note that for $Re\le 135$ the velocity is
negative, i.e.  the spot shrinks, rather than expands.  This regime is difficult
to detect numerically since at these Reynolds numbers the turbulence is not very
stable and can decay spontaneously.  This corresponds to an erosion of the spot
from the inside.  However, inspection of the flow field shows that in the time
interval followed here the velocity given is connected with a retreating front
and not an eroding spot (see also the innermost curves for the smallest Reynolds
numbers in Fig.~9a).  As already mentioned in Sec.~\ref{interior_chap}
the spreading is monotonic in time as
demonstrated by the curves in Fig.~9a.

The
Reynolds number at which the spots start to grow was used by Lundbladh and
Johansson \cite{Lun91} as a definition for the threshold to turbulent behavior.
Here we find that the spot starts to grow for $Re\simeq 135$, a value
compatible with the one determined by the statistical analysis in
section~\ref{sec_basic}, so that the results are consistent with \cite{Lun91}.

\subsection{The propagation mechanism}

By now we have identified three velocities near the boundary of the turbulent
spot:  the front velocity $v_F$ with which the exponentially decaying part of
the envelope propagates, the velocity of the spanwise outflow $U_z$ and the
phase velocity $v_w$ of the streamwise roll pattern.  The value for $U_z$
depends on the distance from the spot, but even when calculated at the position
of the first maximum of the turbulent fluctuations (see e.g.  Fig.~4),
the value is smaller than the front speed $v_F$.  Figure~5 shows that
the outwards travelling roll patterns are eventually overtaken by the turbulent
interior since $v_F$ is larger than $v_w$.

In the absence of a linear instability of the laminar shear flow two
possibilities for the growth mechanism have to be considered:  a linear
instability induced by the cross flow \cite{Gad81,Hen87,Hen89,Dav95} and 
non-normal
amplification \cite{Lan75,Far88,But92,Wal95}.

The combination of basic profile and cross flow defines a rather steady laminar
flow with stability characteristics different from those of the laminar profile.
In the case of plane Poiseuille flow, Henningson could show that the combined
flow is linearly unstable, but the front velocities deduced 
from this instability were
smaller than the observed spreading velocity \cite{Hen87,Hen89}.  A complete
stability analysis would have to include the full profile of the cross flow.  An
estimate of the expected spot growth rates may be based on a local
approximation, where the values of the cross flow are kept fixed.  We thus
determine for each point $z_0$ along the spanwise centerline the cross-flow
\begin{equation}
U_{z}(y,z_0)=
U_{z}(0,z_0)+
\frac{U_{z}(1,z_0)-U_{z}(0,z_0)}{2}[1-\cos(\pi y)]\,,
\end{equation}
where $U_{z}(0,z_0)\simeq U_{z}(2,z_0)$,
and analyze the stability against perturbations
\begin{equation}
v_y(x,y,z,t)=\hat{v}_y(y)\exp[i(k_x x+k_z z-\omega t)]\,,
\end{equation}
where $\omega=\omega_r+i\tilde{\epsilon}$ is a complex frequency.
This leads to the Orr-Sommerfeld equation
\end{multicols}
\begin{eqnarray}
i\omega(D^2-k^2)\hat{v}_{y}+
(ik_x U^{\,''}_{0 x}+ik_z U^{\,''}_{z})\hat{v}_{y}
+Re^{-1}(D^2-k^2)^2\hat{v}_{y}-
(ik_x U_{0 x}+ik_z U_{z})(D^2-k^2)\hat{v}_{y}=0\,,
\end{eqnarray}
\begin{multicols}{2}
with boundary conditions 
\begin{equation}
\hat{v}_{y}(y)=\hat{v}^{\,''}_{y}(y)=0 
\hspace{1em} 
\text{at }\; y= 0 \mbox{\ and \ } 2\,.
\end{equation}
Here, $k^2=k_x^2+k_z^2$.
The primes on the basic profiles $U$ and $D$ denote derivatives with respect to
the wall-normal coordinate $y$.
The local approximation is now reflected in the
fact that the perturbations can have a $z$-dependence, but the basic profile
around which the perturbations are analyzed does not.  With the form of the
cross flow profile and a Fourier expansion for $\hat v_y(y)$ the Orr-Sommerfeld
equation can be solved algebraically.  The maximal growth rates
$\epsilon=\max(\tilde{\epsilon})$ thus obtained are shown in Fig.~10 for
points along the spanwise half-axis and for two values of the Reynolds number.
For the lower value no linear instability is detected.  For Reynolds numbers
$Re\gtrsim 200$ the local growth rate becomes positive, indicating a linear
instability.

Investigations of other front-propagation problems, usually within a
Ginzburg-Landau model, show that it is not only the local instability that
determines the front speed but that the local curvature 
in wave number space has to be included as
well.  In the absence of a derivation of an amplitude equation in a turbulent
medium we phenomenologically take the amplitude in the Ginzburg-Landau equation
to model the envelope of the turbulent intensity $v^2(z,t)$, calculated in
the middle of the cell at $x=L_x/2$ and $y=1$.  The front then connects a
laminar state $(A\simeq 0)$ with a turbulent one $(A\ne 0)$ \cite{Pom91,Saa92}.
For
all practical purposes the turbulent state is stable and the laminar one,
composed of the basic profile and the cross flow, shows a linear instability.
Thus the simplest Ginzburg-Landau model with cubic nonlinearity should be
appropriate, 
\begin{eqnarray}
\partial_t A=\epsilon A+D\partial_z^2 A-b_3\,A^3\,.
\end{eqnarray}
The assumption is that the three parameters $\epsilon$, $D>0$ and $b_3>0$ are
real.  The marginal stability hypothesis then predicts a value for the
asymptotic front velocity of
\begin{eqnarray}
\label{frontv}
v^{\ast}=2\sqrt{\epsilon D}.
\end{eqnarray}
When taking the maximum growth rate $\epsilon$ of our data and evaluating the
diffusion coefficient $D$ by a saddle point approximation around the 
maximum of the
dispersion relation $\omega=\omega(k_x, k_z)$, we end up with a front velocity
which is about an order of magnitude smaller than the observed one. Corresponding to
Eq.~(\ref{frontv}), one gets, e.g., $v^{\ast}\simeq 0.06$ for $\epsilon\simeq 0.017$,
$D\simeq 0.056$, and $Re=300$ (see Fig.~10).
However, at this Reynolds number the front moves with $v_F\simeq 0.63$.
\begin{figure}
\begin{center}
\epsfig{file=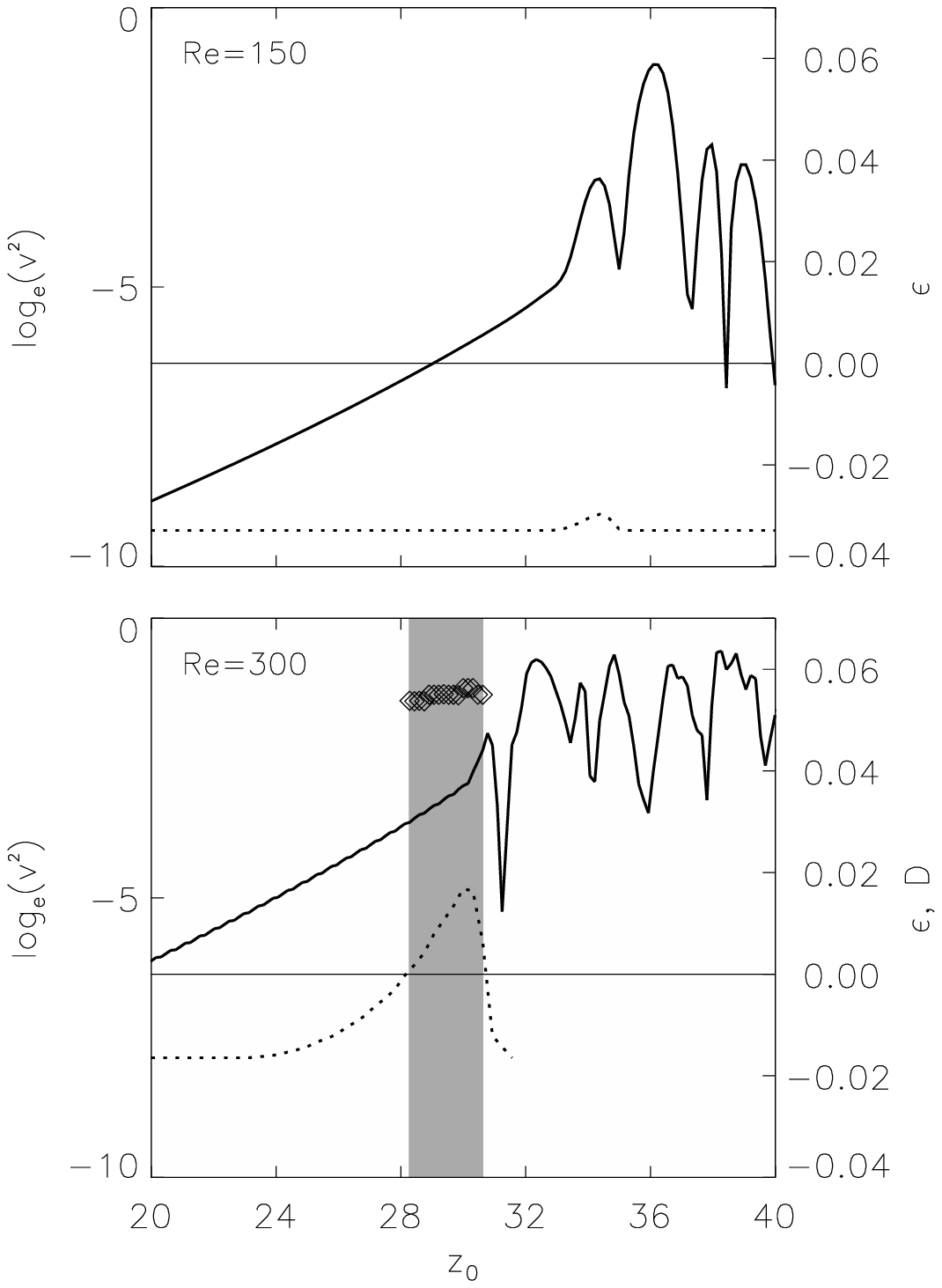,width=6cm}
\end{center}
\end{figure}

\noindent
FIG. 10. Maximum growth rates $\epsilon$ of plane waves perturbing the
mean flow combined with the cross-stream outflow along the spanwise axis.
The sets for two different Reynolds numbers are compared at $t=39$. 
Thick solid lines denote $\log_e(v^2)$ and thick dotted lines denote
maximum growth rate $\epsilon$. Additionally, for $Re=300$ the diffusion
coefficient $D$, which follows from the dispersion relation, is plotted for
the gray shaded range of $z_0$ values.

\vspace{0.5cm}
\noindent
This analysis of the front propagation mechanism, with a local approximation,
only cubic terms in the amplitude model and without a discussion of slow
transients \cite{Ebe00} that could arise in such models, is rather simplified
and can only give an indication of the expected front velocity.  Nevertheless,
it seems to us that even when these improvements are included the model cannot
account for the observed spot spreading rates:  first of all, it cannot say
anything about the dynamics below a Reynolds number of about 200, where the spot
spreads but where there is no instability.  And secondly, even above this
Reynolds number, where there is a linear instability, the calculated front speed
differs significantly from the observed one.

Besides the linear instability mode for spot growth there is another possibilty,
based on the non-normal amplification of perturbations near the spot interface.
Because of the action of the perturbation on the basic profile, streamwise
vortices need not decay monotonically but can first grow on a time scale of
order $Re$ to an amplitude about a factor $Re$ larger than the initial
amplitude.  This so-called lift-up effect \cite{Lan75}
is most likely responsible for the
occurence of streamwise streaks in turbulent shear flows where longitudinal
modulations can give rise to secondary instabilities and a perpetual
non-periodic time evolution \cite{Gro99,Wal95}.  Indeed we do observe this
cyclic reproduction of the coherent structures in the interface region of the
spot as demonstrated in the sequence of Fig.~5.  The model then is that
the turbulent interior of the spot induces a small perturbation near the
interface which will then be amplified and grow turbulent.  The statistical
analysis of section~\ref{sec_basic} shows that not all perturbations grow
turbulent.  Thus, if most of the perturbations grow turbulent, the spot will
spread, but if most of the perturbations decay, it will shrink.  The observed
coincidence between the critical Reynolds number for transition and the one for
spot growth can naturally be explained in this picture.  However, we do not see
how to derive other quantitative conclusions from this model.  In particular,
the front speed would be given as the quotient of the width of the rolls
generated (this can be read off rather accurately from the frames) and the time a
perturbation needs to grow turbulent.  The latter depends on the amplitude of
the initial seed, the threshold for the transition to turbulence and the
amplification rate, neither of which seems accessible to independent
determination.

\section{Final remarks}
\label{sec_summary}

Our analysis of a shear flow with free slip boundary conditions on two parallel
surfaces has revealed many similarities to plane Couette flow between rigid
walls.  As in that case three velocities connected with the spot can be
identified, the velocities of the advancing front, of the outward flow component
and of the phase speed of (oblique) waves.  They differ in value and in
$Re$-dependence, and relations between them are unknown.  The dependence of the
front velocity with respect to Reynolds number is consistent with experimental
findings \cite{Til92,Dav95}.  Many of the results reported here parallel the
ones for plane Poiseuille flow.  We also find waves and instabilities in the
neighborhood of the spot, but they do not lead to quantitative predictions for
the front velocity.

The conclusions we draw from this investigation highlight a dilemma:  On the one
hand side the large scale flow outside the spot does not seem to be important:
the spot grows independent of whether there is a linear instability of basic
flow plus large scale exterior flow or not, and if there is a linear instability
the derived front speed is slower than the observed one.  Moreover, the
behavior expected from the non-normal amplification mechanism can explain some
aspects of the dynamics, especially for lower Reynolds numbers.  On the other
hand, if the outflow is suppressed, as in the case of a turbulent ribbon that
spans the cell in streamwise direction, no growth is observed
despite the random initialization of seeds near the spot interface.  Perhaps this can
explain the observed stop of growth in some experiments \cite{Dav95}.  So it
seems that spot growth is a subtle interplay between local features (e.g.
non-normal amplification) and global features (such as the external flow).  The
connection between both and a quantitative estimate of the growth velocity
remain a major puzzle in the dynamics of turbulent spots in parallel shear
flows.

\acknowledgments 
We thank Paul Manneville for discussions and 
the John-von-Neumann Institut f\"ur Computing in J\"ulich for
computing time on a Cray T-90 without which this study would not have
been possible.


\end{multicols}

\begin{thebibliography}{10}
\bibitem{Rey83} O. Reynolds, Proc. R. Soc. Lond. {\bf 35}, 84 (1883).

\bibitem{Cha73} I. J. Wygnanski and F. H. Champagne, J. Fluid Mech. {\bf 59},
		    281 (1973).
		    
\bibitem{Col65} D. Coles, J. Fluid Mech. {\bf 21}, 385 (1965).
		    
\bibitem{Heg89} J. J. Hegseth, C. D. Andereck, F. Hayot, and Y. Pomeau,
		    Phys. Rev. Lett. {\bf 62}, 257 (1989). 

\bibitem{Til92} N. Tillmark and P. H. Alfredsson, J. Fluid Mech. {\bf 235},
		    89 (1992).
		    
\bibitem{Dav92} F. Daviaud, J. J. Hegseth, and P. Berg\'{e}, Phys. Rev. Lett.
		    {\bf 69}, 2511 (1992).

\bibitem{Bo98}  S. Bottin, F. Daviaud, P. Manneville and O. Dauchot,
		    Europhys. Lett. {\bf 43}, 171 (1998).
		    
\bibitem{Man00}  P. Manneville and O. Dauchot, {\em Patterning and 
                 transition to turbulence in subcritical systems: the case
                 of plane Couette flow}, ed. M. Rubi, to be published in 
                 Proceedings of XVII Sitges conference.

\bibitem{Gad81} M. Gad-El-Hak, R. F. Blackwelder, and J. J. Riley,
		    J. Fluid Mech. {\bf 110}, 73 (1981).                               

\bibitem{Cro92} M. C. Cross and P. C. Hohenberg, Rev. Mod. Phys. {\bf 65}, 851
		 (1992).

\bibitem{Pom91} Y. Pomeau, Physica D {\bf 23}, 1 (1986); {\bf 51}, 546 (1991).
		    
\bibitem{Cro80} M. C. Cross, Phys. Fluids {\bf 23}, 1727 (1980).
		    
\bibitem{Saa92} W. van Saarloos and P. C. Hohenberg, Physica D {\bf 56},
		    303 (1992).
		    
\bibitem{Schm97} A. Schmiegel and B. Eckhardt, Phys. Rev. Lett. {\bf 79},
		     5250 (1997).
								
\bibitem{Schm99} A. Schmiegel and B. Eckhardt, Europhys. Lett. {\bf 51}, 395
                 (2000).
		     
\bibitem{Gro99} S. Grossmann, Rev. Mod. Phys. {\bf 72}, 603 (2000). 
		    
\bibitem{Hen87} D. S. Henningson and P. H. Alfredsson, J. Fluid Mech. {\bf 178},
		    405 (1987).

\bibitem{Hen89} D. S. Henningson, Phys. Fluids A {\bf 1}, 1876 (1989).
		    		    
\bibitem{Lun91} A. Lundbladh and A. V. Johansson, J. Fluid Mech. {\bf 229},
		    499 (1991).
		    
\bibitem{Lun93} A. Lundbladh, {\it Simulation of Bypass Transition to 
		    Turbulence in Wall Bounded Shear Flows}, PhD thesis,
		    Royal Institute of Technology, Stockholm (1993).

\bibitem{Til95} N. Tillmark, Europhys. Lett. {\bf 32}, 481 (1995). 
		    
\bibitem{Dav95} O. Dauchot and F. Daviaud, Phys. Fluids {\bf 7}, 342 (1995).

\bibitem{Mal95} S. Malerud, K. J. M{\aa}l{\o}y, and W. I. Goldburg, 
                Phys. Fluids {\bf 7}, 1949 (1995).

\bibitem{Heg96} J. J. Hegseth, Phys. Rev. E {\bf 54}, 4915 (1996). 

\bibitem{Dra81} P. G. Drazin and W. H. Reid, {\em Hydrodynamic Stability},
		    (Cambridge University Press, Cambridge, 1981). 
		    
\bibitem{Can88} C. Canuto, M. Y. Hussaini, A. Quaternioni, and T. A. Zang,
		    {\em Spectral Methods in Fluid Dynamics},
		    (Springer, Berlin, 1988). 
		    
\bibitem{See96} N. Seehafer, E. Zienicke, and F. Feudel, 
		    Phys. Rev. E {\bf 54},  2863  (1996).

\bibitem{Wal97} F. Waleffe, Phys. Fluids {\bf 9}, 883 (1997).
		    
\bibitem{Schm99a}  A. Schmiegel and B. Eckhardt, {\em Dynamics of perturbations
		in plane Couette flow}, (unpublished).

\bibitem{Lan75} M. T. Landahl, SIAM J. Appl. Math. {\bf 28}, 735 (1975).

\bibitem{Far88} B. F. Farrell, Phys. Fluids {\bf 31}, 2093 (1988).

\bibitem{But92} K. M. Butler and B. F. Farrell, Phys. Fluids A {\bf 4},
                1637 (1992).

\bibitem{Wal95} F. Waleffe, Stud. Appl. Math. {\bf 95}, 319 (1995).
                
\bibitem{Ebe00} U. Ebert and W. van Saarloos, Physica D {\bf 146}, 1 (2000).
                
\end{thebibliography}
\end{document}